\documentclass[oneeqnum,onefignum,letterpaper,final]{siamltex1213}
\usepackage{amsmath}
\usepackage{amssymb}
\usepackage{hyperref}

\newcommand{\Ubar}{\ensuremath{\widehat{U}}}
\newcommand{\Vbar}{\ensuremath{\widehat{V}}}
\newcommand{\Npv}{\ensuremath{n_{\text{PV}}}}
\newcommand{\Npq}{\ensuremath{n_{\text{PQ}}}}
\newcommand{\splus}{\ensuremath{s_\text{+}}}
\newcommand{\sminus}{\ensuremath{s_\text{-}}}

\title{FUNDAMENTALS OF THE HOLOMORPHIC EMBEDDING LOAD-FLOW METHOD}
\author{Antonio Trias\thanks{Aplicaciones en Inform\'atica Avanzada, Sant
    Cugat del Vall\`es, 08172 Spain (\texttt{triast@aia.es}).}}

\begin{document}

\maketitle 

\begin{abstract}
  The Holomorphic Embedding Load-Flow Method (HELM) was recently introduced as
  a novel technique to constructively solve the power-flow equations in power
  grids, based on advanced complex analysis. In this paper, the theoretical
  foundations of the method are established in detail. Starting from a
  fundamental projective invariance of the power-flow equations, it is shown
  how to devise holomorphicity-preserving embeddings that ultimately allow
  regarding the power-flow problem as essentially a study in algebraic curves.
  Complementing this algebraic-geometric viewpoint, which lays the foundation
  of the method, it is shown how to apply standard analytic techniques (power
  series) for practical computation. Stahl's theorem on the maximality of the
  analytic continuation provided by Pad\'e approximants then ensures the
  completeness of the method. On the other hand, it is shown how to extend the
  method to accommodate smooth controls, such as the ubiquitous
  generator-controlled PV bus.
\end{abstract}



\pagestyle{myheadings}
\thispagestyle{plain}
\markboth{A.~TRIAS}{FUNDAMENTALS OF HELM}

\section{Introduction}

The power grid has been hailed by the US National Academy of Engineering as
the most influential engineering innovation of the 20th
century~\cite{constable2003century}. Electrical power is what makes modern
society tick, and the power grid has become a critical infrastructure. It is
essentially a network of high voltage lines, transformers, and substations
that carries bulk power over long distances, from power generation facilities
to distribution substations.

The cornerstone problem of AC electrical power systems is the so-called
\emph{power-flow study} (also known as \emph{load-flow}), which describes the
steady state of the network under some given conditions. The problem can be
written as follows, in terms of the current balance at each bus $i$:
\begin{equation}
  \sum_k {Y_{ik}^\text{(tr)} V_k } + Y_{i} ^\text{(sh)} V_i =
  \frac{S_i^*}{V_i^*}
  \label{eq:powerflow_study}
\end{equation}
where $Y_{ik} ^\text{(tr)}$ are the elements of the transmission admittance
matrix, $Y_{i} ^\text{(sh)}$ are shunt admittances, and $S_i$ are
constant-power injections going into the bus. The index $k$ runs over all
buses including the swing bus, whose voltage $V^\text{(sw)}$ is specified as
the reference.  In its most basic form, the problem consists in
solving~\eqref{eq:powerflow_study} for the voltages $V_i$ (for all $i$ except
the swing), for a given set of injections $S_i$. A variation of this problem,
closer to actual practice, involves contemplating so-called PV buses, in which
the voltage magnitude $|V_i|$ is kept constant by means of a variable
injection of reactive power $Q_i$ by some generator (this amounts to adding
new constraint equations and some corresponding new variables
to~\eqref{eq:powerflow_study}, as it will be shown in
\S~\ref{sec:extensions}).  In any case, note that the l.h.s.\ terms are all
linear, but the constant-power injections appearing on the r.h.s.\ make the
problem non-linear and multi-valued in general.

Many numerical methods have been devised for solving this problem since the
beginning of computing. The earliest ones were based on Gauss-Seidel (GS)
iteration~\cite{WardHale56}, which has slow convergence rates but very small
memory requirements. Most other methods are based on Newton--Raphson
(NR)~\cite{TinneyHart67}, which is generally better than GS because of its
quadratic convergence properties.

Several improvements on the standard NR method were developed by exploiting
the weak coupling between reactive power and voltage magnitude on the one
hand, and real power and phase angles on the other, which yields good
approximations in high-voltage transmission systems. Decoupling leads to
smaller Jacobian matrices, which is a big computational gain in large
networks, even when using sparse linear algebra techniques for maximum
efficiency.  Of all the various decoupled methods based on NR, the so-called
Fast Decoupled Load Flow (FDLF) formulation of Stott and
Alsac~\cite{StottAlsac74} has become the most successful and it is almost a
de-facto standard in the industry, either in its original form or in one of
its variants~\cite{Amerongen89}. In addition to decoupling real and reactive
power, the FDLF method factorizes the Jacobian matrix only once. Standard
textbooks on power system
analysis~\cite{grainger1994power,kundur1994power,das2011power} describe all
these methods in detail, and there exist widely available, open-source
reference implementations of them, such as MATPOWER~\cite{MATPOWER}.

A common shortcoming of these traditional methods is their reliance on
numerical iteration as the root-finding technique lying at the core of the
procedure.  As it is well-known, the success of NR or similar contraction-map
iterations depends on the choice of an initial seed.  Indeed, several authors
have verified that since the powerflow problem is multi-valued, iterative
methods not only exhibit this dependency, but may also behave erratically,
since the different basins of attraction for the various solutions intertwine
along fractal
boundaries~\cite{ThorpNaqavi97,Mori00,KlumpOverbye00b}. Certainly these
problems may be overcome in the vast majority of cases if one devotes some
time to explore different starting seeds and monitors the result to avoid
undesired solutions. Additionally, many authors have developed additional
techniques to improve the chance of convergence or even ensure global
convergence properties~\cite{AjjarapuChristie92,KlumpOverbye00a,Mehta14}, but
their computational cost is significantly higher, their convergence is still
not completely ensured in all cases, and human supervision is still needed to
assess the different solutions obtained.

The Holomorphic Embedding Load-flow Method (HELM) method~\cite{Trias12} was
born out of the necessity to have a powerflow solution method that could run
completely unattended and still produce reliable solutions. These
characteristics are absolutely needed for building modern applications that
perform massive searches in the state-space of the system, such as intelligent
decision-support systems for grid operators.  To give a simpler example,
consider computing the solution of the system under the outage of one or more
lines, which is a very common security assessment procedure. In that scenario
NR-based methods have a non-negligible probability of failing, since the new
solution may be ``too far'' from the previous undisturbed state.

From a numerical methods point of view, HELM is direct and constructive, and
it does not require the choice of an initial seed. Based on a holomorphic
embedding technique, it allows computing the formal power series
corresponding to the desired solution in an unequivocal way. Then, thanks to a
series of recent advances in the theory of Pad\'e approximants, the numerical
solution can be obtained with maximal guarantees (within the limits of
floating point precision and round-off). The method thus obtains the desired
solution if it exists, and conversely it unequivocally signals unfeasibility
when such solution does not exist. It should be remarked that this method has
been successfully implemented commercially, and proven able to solve networks
of over 65,000 buses with performance that is competitive with traditional
fast-decoupled methods.

A couple of clarifications regarding seemingly related methods is needed
here. Although the idea of embedding is also used in so-called continuation
powerflow methods~\cite{AjjarapuChristie92}, note that HELM is based on
\emph{holomorphicity}, which is a much more strict requirement than the simple
smoothness properties required by homotopy. Moreover, as shown in this
paper, the underlying formulation in terms of algebraic curves provides a
global view that is harder to obtain otherwise.  On the other hand,
other authors have also explored the idea of exploiting the smoothness of the
problem through the use of Taylor series expansions of the real and imaginary
parts of the variables (or equivalently, their magnitudes and
angles)~\cite{Sauer81,Xu98a,Zambroni07}. This is of course very different from
exploiting holomorphicity and analytic continuation.

In this paper, the theoretical foundations of the HELM method presented
in~\cite{Trias12} are established in detail. The emphasis is on algebraic
curves, showing how two basic conceptual components of that field, namely
polynomials and power series, apply to the powerflow problem. Additionally, it
is shown how the method can accommodate the inclusion of any type of smooth
control within exactly the same framework. The paper is structured as follows.
Section~\ref{sec:overview} provides a quick overview of the method,
summarizing the exposition given in~\cite{Trias12} from a conceptual point of
view.  Section~\ref{sec:fundamentals1} discusses the foundations of the method
by establishing the underlying mathematical structure of the theory, based on
algebraic curves. This gives new insights into the powerflow problem.
Section~\ref{sec:fundamentals2} provides the fundamentals for the more
pragmatic aspect of the method, i.e. computing the solution through power
series and analytical continuation.  Then \S~\ref{sec:extensions} shows
how the method can accomodate powerflow controls in a natural way.  Some
illustrative and pedagogical examples are fully worked out in the Appendix.

\section{\label{sec:overview} Overview of HELM}

Our intention here is to show that the holomorphic embedding method is in fact
a procedure to make the powerflow interpretable within the general framework
of algebraic curves. This can be considered as looking at the problem from the
viewpoint of polynomial systems. Doing this allows us to bring to the
powerflow problem the amazing concepts of algebraic geometry, a subject that
marries advanced complex analysis and abstract algebra, and dubbed by many
authors as ``the jewel'' of XIX century mathematics. To do so, one needs to
embed the complex voltage variables as holomorphic functions. For this it is
necessary to avoid the use of $V_i^*(s)$, which is not holomorphic, and use
instead the variable $\Vbar_i(s)$ as an independent holomorphic function. The
original problem is recovered at the evaluation on the focal point by imposing
the ``reflection'' condition $\Vbar_i(s)=V_i^*(s^*)$.

Let us begin by reviewing the steps of the method in outline mode, and expand
on their meaning later on. The first steps adopt the algebraic curve
(polynomial) viewpoint, laying down the foundational concepts:
\begin{enumerate}
\item Embed the equations using a complex parameter $s$ as shown
  in~\cite{Trias12}. To obtain holomorphicity in the next step, it is crucial
  that the variables $V_i^*$ are embedded as $V_i^*(s^*)$, \emph{not} as
  $V_i^*(s)$.
\item Use $\Vbar_i(s)$ instead of $V_i^*(s^*)$; now the embedded equations
  define an algebraic curve and therefore the variables are holomorphic
  functions. However, once solved, we have to remember to request the
  reflection condition at $s=1$. Only such solutions represent physical
  (feasible) branches of the original powerflow problem; the rest are ghost
  solutions. \item The system of equations are polynomial. Gr\"obner basis
  theory demonstrates that, using elimination techniques (for instance with
  lexicographical ordering), it is possible to arrive at a single polynomial
  equation in just one of the variables, say $V_1(s)$, while the rest,
  including of course $\Vbar_k(s)$, are recovered univocally in a triangular
  fashion. A single polynomial equation with $s$-dependent coefficients is by
  definition a \emph{plane algebraic curve}.
\item Problems of existence of non ghost solutions (both operational and
  non-operational) can now be studied in terms of the topology, singularities,
  and branching points of the algebraic curve. For many small models, one may
  obtain illuminating results in closed form. This is not only useful for
  instructional analysis, but also for the exact study of small network
  equivalents, in particular the network algebraic equivalents that naturally
  arise in HELM, inspired in the theory of algebraic curves (these will be
  discussed in a forthcoming paper).
\end{enumerate}

The rest of the steps deal with the computational aspects. The viewpoint thus
switches from polynomials to power series that describe the local behaviour of
algebraic curves at every point:
\begin{enumerate}
\setcounter{enumi}{4}
\item As the number of variables grow, it is not computationally feasible to
  use polynomial elimination techniques to solve for the algebraic curve
  explicitly. Instead, the method calculates the power series of the curve at
  the reference point $s=0$ (for a suitably chosen germ, which defines the
  branch), and then uses analytic continuation to try to reach the target
  point, $s=1$. The fact that one is dealing with an algebraic curve ensures
  that this series exists and has a non vanishing convergence radius.
\item The germ of choice at $s=0$ is the one that physically corresponds to an
  energized network with no constant-power loads or injections. It verifies
  the reflection condition.
\item The analytic continuation is computed by near-diagonal sequences of Pad\'e
  approximants of the power series. Stahl's theorem guarantees that the result
  is single-valued and maximal: if the approximants converge at $s=1$, we have
  obtained the solution; otherwise, there is no feasible powerflow.
\end{enumerate}

Note that, under this procedure, the choice of a particular solution branch at
$s=0$ uniquely determines the powerflow solution that is obtained at $s=1$, if
it exists. The choice that the method proposes, which actually defines what we
refer to as the \emph{operational} solution, is easily identified as the state
with zero constant-power load/generation \emph{and} non-zero voltage
throughout the network (i.e.\ zero power because the bus is open-circuited;
note that zero power could also be achieved by short-circuiting, which gives
rise to ``dark'' branches~\cite{Trias12}).

This has been a summarized overview of the logic steps in building the
theoretical scaffolding underlying the method. An actual numerical
implementation would only deal with the calculation of power series and their
Pad\'e approximants~\cite{BakerGraves96}, but such narrow mechanistic view of
the method, although perfectly viable as a numeric procedure, would ignore the
wealth of insights coming from the global aspects of the algebraic curve. We
now leave the outline style and expand upon the motivation and foundations of
each of these stages of the method.

\section{\label{sec:fundamentals1} Fundamentals I: the algebraic view}

\subsection{Preliminaries}
The powerflow problem will be assumed to have the following general form:
\begin{equation}
  \sum_k {Y_{ik} V_k } = \frac{S_i^*}{V_i^*}
  \label{eq:LF}
\end{equation}
where $Y_{ik}$ is the generalized admittance matrix containing the effects of
all linear devices. This typically includes line admittances, transformers,
(including phase-shifters), line susceptances, bus shunt admittances,
constant-impedance loads, and constant-current injections, among others. The
r.h.s.\ contains the constant-power injections, which are the nonlinear part.

Unless specified, indices run over all buses, including the swing
bus which will be chosen at index $0$.

\subsection{\label{subsec:motivation} Motivation}

The major motivation driving the methodology behind HELM originates in the
\emph{projective invariance} of equations~\eqref{eq:LF}, which strongly
suggests how to study the system. If the voltages are rescaled by
$V'_i=\lambda V_i$, the resulting equations recover the same form, but with
scaled injections:
\begin{equation}
  \sum_k {Y_{ik} V'_k } = |\lambda|^2 \frac{S_i^*}{{V'}_i^*}
 \label{eq:proj_invar}
\end{equation}
As it is well-known in other areas of physics, this sort of invariance is
something that deserves to be studied in its own right, as it often rewards us
with insights and powerful results that would otherwise be lost (e.g., gauge
theories in theoretical physics). In this case the projective invariance
in~\eqref{eq:proj_invar} can be understood as a scale invariance linking the
scales of voltages and power injections, so that the equations describe a
whole family of powerflow problems at a time. Naturally, one fixes this
invariance when choosing a particular reference value of the swing voltage,
but the point here is that we would like to study this dependence. Therefore
this leads one to consider an embedding technique by using $s=|\lambda|^2$ as
a new variable of the problem. The family corresponds to all real and positive
values of $s$.

However, there are strong reasons to work with a complex embedding parameter.
The first reason is that equations~\eqref{eq:LF} are algebraic, and therefore
they can always be reduced to polynomials by means of elimination techniques,
as shown in \S~\ref{subsec:elimination}. Therefore it makes sense to work
in the complex domain, where the fundamental theorem of algebra guarantees
that all zeros exist. Another reason is that, using a suitable form of the
embedding as shown below, the problem is converted into an algebraic
curve. This opens up a plethora of techniques and results from complex
analysis. Most of them derive from \emph{holomorphicity}, i.e.\ complex
analyticity of the voltages with respect to the complex embedding
parameter. Therefore there is nothing to lose and a lot to win by working in
the complex $s$-plane (eventually, the functions are evaluated at the focal
real value of $s$ to get the desired voltages). In fact the main proposition
in HELM is to look at the powerflow as a particular problem within the theory
of algebraic curves. This requires considering complex variables and
holomorphic functions.

\subsection{Complex embedding}

The proposed embedding consists in introducing a complex parameter $s$ into
(\ref{eq:LF}). A natural possibility, which will be referred to as the minimal
embedding, is the one suggested by the projective invariance discussed above:
\begin{equation}
   \sum_k {Y_{ik} V_k(s)} = s \frac{S_i^*}{V_i^*(s^*)}
   \label{eq:HolEmb}
\end{equation}
Note how one recovers exactly the powerflow equations of~\eqref{eq:proj_invar}
for real, positive values of $s$. On the other hand, it is key that the
voltage parameters $V_i^*$ are embedded as $V_i^*(s^*)$, \emph{not} as
$V_i^*(s)$, since the former verifies the Cauchy-Riemann equations and the
latter does not. The idea is also inspired by Schwarz reflection
principle~\cite{Marsden99}, and its use will become clear in the next section.

\subsection{Holomorphicity and the reflection condition on $V, \Vbar$}

Since complex conjugation does not preserve holomorphicity, the embedded
system~(\ref{eq:HolEmb}) must be formally doubled with its mirror image:
\begin{equation}
  \begin{split}
    \sum_k{Y_{ik} V_k(s)} &= s \frac{S_i^*}{\Vbar_i(s)} \\
    \sum_k{Y_{ik}^* \Vbar_k(s)} &= s \frac{S_i}{V_i(s)}
  \end{split}
  \label{eq:HolEmbFull}
\end{equation}
where $V_i(s),\Vbar_i(s)$ must be considered as two sets of independent
holomorphic functions. Note that here the hat denotes just a different
symbol, not complex conjugation. Once the solution for both sets is found,
if the following equality holds
\begin{equation}
  \Vbar_i(s) = V_i^*(s^*),
  \label{eq:reflexcond}
\end{equation} 
which will be referred to as the \emph{reflection condition}, then the two
sets of equations in~(\ref{eq:HolEmbFull}) are just complex conjugates of each
other, and the embedded system~(\ref{eq:HolEmb}) is recovered. Note however
that the doubled system~(\ref{eq:HolEmbFull}) may contain solutions that do
not satisfy the reflection condition~(\ref{eq:reflexcond}) when evaluated at
$s=1$ and therefore are not physical solutions to the original powerflow
problem (these will be referred to as ghost solutions). Therefore the method
consists in solving the algebraic system~(\ref{eq:HolEmbFull}) \emph{and}
requiring the additional condition~(\ref{eq:reflexcond}) to hold at
$s=1$. Then, among the non-ghost solutions, if any, one will identify the
one representing the normal operational condition as a well-defined specific
branch of the algebraic curve (white branch).  All other non-ghost solutions
are black branches, i.e.\ physical in principle but representing low-voltage,
non-operational scenarios. If all the algebraic solutions are ghost, the
original parametric powerflow problem has no solution.

\subsection{\label{subsec:elimination}Elimination polynomial}

Eliminating denominators, the system~(\ref{eq:HolEmbFull}) becomes a set of
polynomial equations in several variables. There exist many elimination
techniques to solve these~\cite{Sturmfels02}, such as classical resultants,
characteristic sets, and Gr\"obner bases.  Although in practice the HELM
method does not use any of these computer algebra methods, Buchberger's
algorithm on Gr\"obner basis theory~\cite{Buchberger98} may be invoked to
prove that it is always possible to carry out a complete elimination
procedure, arriving at a polynomial equation in only one of the variables
(say, $V_1$):
\begin{equation}
  \mathfrak{P}(s, V_1) = \sum_{n=0}^{\deg\mathfrak{P}} {a_n(s)V_1^n} = 0
  \label{eq:PolyEq}
\end{equation}
Furthermore, using lexicographical monomial ordering, the elimination is
triangular: all the other variables $\Vbar_1,V_2,\Vbar_2,V_3$, etc., are
expressed explicitly as polynomials in the previous ones. Given a solution
$V_1(s)$ to~(\ref{eq:PolyEq}), all other values $V_i(s), \Vbar_i(s)$ are then
obtained by simple progressive back-substitution.

The degree $\deg\mathfrak{P}$ of this polynomial is in general rather large
(of order exponential in the number of buses), but the key points here are
that the degree is always finite and the coefficients $a_n(s)$ are polynomial
in $s$. This means that~\eqref{eq:PolyEq} is a \emph{plane algebraic curve}.
Thus the powerflow problem has been converted, via the embedding procedure,
into the study of algebraic curves on the complex plane, which is a field in
which there is a plethora of results to exploit and build
upon~\cite{Fischer2001,Kunz2007}. Algebraic curves are essentially linked to
Riemann surfaces~\cite{Springer2001,Farkas2012}, another field rich in
powerful results.

\subsection{Algebraic curves. Branches}

One of the most immediate results that can be reaped concerns the analysis and
interpretation of the multiple solutions of the powerflow problem, and the
collisions among them, in particular at voltage collapse points.  Now all
possible powerflow solutions are characterized as \emph{branches} of the
corresponding algebraic curve defined by~(\ref{eq:PolyEq}).  Branch collisions
take place at the so-called \emph{branch points} of the curve, which are the
values of $s$ for which a zero of $\mathfrak{P}(V_1)$ has multiplicity 2 or
greater.  This, together with the reflection condition~(\ref{eq:reflexcond}),
allows us to interpret feasibility and voltage collapse in the powerflow
problem under a new light. Algebraic curves provide a global view with far
greater analytical power than numerical techniques such as Newton--Raphson,
which only exploit local properties. For instance, a traditional continuation
powerflow~\cite{AjjarapuChristie92} can calculate and analyze the collapse
point in terms of a saddle-node bifurcation, but the theory presented here
reveals it more specifically as two branches of a (complex) algebraic curve
merging at a branch point.

At each particular point $s$ of the complex plane, the branches can be
categorized into these groups:
\begin{itemize}
\item Ghost branches: these correspond to solutions that fail to satisfy the
  reflection condition. Therefore, when $s$ is real positive, they are not
  physical powerflow solutions.
\item Feasible branches: these satisfy the reflection condition and therefore
  represent physically possible powerflows when $s$ is real positive. However,
  only one of these corresponds to the normal operating state of a power
  network, as it will be shown below.
\end{itemize}
Voltage collapse is therefore understood as a collision in the $s$-plane of the
operational branch with another one and the emergence of a couple of ghost
branches as a result.

Algebraic techniques allow the calculation of all branches simply as the roots
of polynomial $\mathfrak{P}(s, V_1)$. Calculation of all branch points is also
accomplished by well-known techniques, such as the calculation of the
discriminant of $\mathfrak{P}(s, V_1)$ (the resultant of the two polynomials
$\mathfrak{P}, \partial \mathfrak{P}/\partial V_1$). Appendix~\ref{app:twobus}
demonstrates the power of this analysis in an example where the curve and its
branch points can be explicitly calculated. However, since the degree of the
algebraic curve grows exponentially with the number of buses, such explicit
calculations can only be carried out for very small networks.  Several
authors~\cite{MontesCastro95,Montes98,KavasseriNag08,Ning09,NguyenTuritsyn14}
have explored these computer algebra techniques for simultaneously obtaining
all solutions to the powerflow problem (although not in an embedded setting),
and the typical size that is computationally feasible remains at around four
to five buses maximum.  Therefore the exposition turns now to the method for
computing solutions for networks of any size.

\section{\label{sec:fundamentals2} Fundamentals II: the analytical view}

\subsection{Power series}

As it is customary in the field of algebraic curves, for practical
computations one uses the power series representation developed about some
reference point. Since the $V_i(s)$ are holomorphic, their power series
contain all the information needed to reconstruct the functions in their full
domain of holomorphy (which in this case, being algebraic curves, is the full
complex plane except a finite number of singular points), beyond the
convergence radius of the series. Each power series defines a different branch
and in this sense, they are referred to as ``germs'' of the branches of the
algebraic curve~\cite{Ahlfors79}. This full reconstruction is provided by the
powerful Weierstrass \emph{analytic continuation} procedure, and it is what
allows HELM to extend the germs from local to global branches.

In sum, working with power series allows practical computation in networks of
any size, and holomorphicity ensures that, at least in principle, calculations
at points far away from the reference can be carried out through analytic
continuation.

\subsection{The choice of branch. Operational solution}

The next problem consists in selecting the proper branch. Since the goal is to
obtain the operational solution of the original powerflow~(\ref{eq:LF}), some
criterion is needed to select a branch out of the multiple ones in
system~(\ref{eq:HolEmbFull}). Drawing from the physics behind the problem, it
can be argued that $s=0$ provides the privileged reference point for an
unambiguous choice. At $s=0$ the system represents a power network in which
the constant-power injections (whether load or generation) are zero. At each
bus, this situation can be achieved either by open-circuit, in which case the
voltage is non-zero, or by short-circuit, in which case the voltage is zero
and the right-hand side $sS_i^*/\Vbar_i(s)$ in~(\ref{eq:HolEmbFull})
approaches the value of the bus fault current. Requiring that $V_i(0)\neq 0$
throughout, i.e.\ open circuiting all buses except the swing, provides a
unique answer, which will be referred to as the white branch (at $s=0$).

Reference~\cite{Trias12} shows in detail the mechanics to compute the power
series terms of this white germ, up to any desired order, through a
constructive procedure involving the repeated solution of a linear system in
sequence. One important point is that this procedure automatically enforces
the reflection condition; the white germ clearly satisfies this condition at
$s=0$.  Regarding a power series as the identity mark of a branch, and
Weierstrass analytic continuation as a propagation of this identity to other
parts of the $s$-plane, HELM postulates that the operational solution sought
at $s=1$ must have the same identity and therefore is the analytic
continuation of the white germ.

\subsection{Analytic continuation and single-valuedness}

In HELM one assumes that the powerflow problem corresponds to the operational
state of some real network. Therefore, regardless of the particular method
used to perform the analytic continuation of the white germ discussed above,
we need to enforce the single-valuedness of the procedure in order to obtain a
single solution. It is well-known that analytic continuation of a holomorphic
function from a given point $s_a$ to a point $s_b$ along different paths may
yield different values of the function at $s_b$ just by following different
paths. This reflects the fact that the algebraic curve is actually one whole
multi-valued holomorphic function, not simply a collection of disjoint
branches. This ``branch identity loss'' can only happen if both paths enclose
a singular point (branching point).

The way to solve this issue is standard in complex analysis. One must remove
from the complex plane (technically, from the extended complex plane
$\mathbb{C}_\infty$, the Riemann sphere) a connected set of lines $\Gamma$
(branch cuts) connecting all branch points. Then the complement
$\mathbb{C}_\infty\setminus\Gamma$ will be simply connected (no holes) and the
analytic continuation along any path on the remaining plane will be guaranteed
to provide a unique result, since it is no longer possible to have paths
encircling a branch point (because they would have to cross one of the removed
lines). This is for instance what is done in the definition of some elementary
multi-valued complex functions, such as the square root, in which the choice
of the branch cut along $(-\infty,0)$ fixes the conventional meaning of the
principal value of the square root (i.e.\ the meaning of the symbol $\sqrt z$).

Note that the choice of branch cuts is a priori arbitrary (as long as they
connect all branch points), but every choice leads in principle to a different
single-valued function. Therefore the problem is not only achieving
single-valuedness, but also choosing the set $\Gamma$ (branch cuts) in such a
way that the analytical continuation of the white germ at $s=1$ gives a
solution that makes physical sense. Our argument, based on physical
plausibility, is that \emph{the operational solution to the powerflow problem
  exists if and only if, using the complex embedding~\eqref{eq:HolEmb}, the
  white germ is analytic-continuable along all values $0<s\le1$ on the real
  axis}. Therefore the only a priori requirement on the choice of branch cuts
is that they do not contain any part of this particular continuation path.
Intuitively, one requires that the whole family of projectively equivalent
powerflow problems up to $s=1$ retains the same identity as the germ at $s=0$.

\subsection{Maximality of the analytic continuation: Stahl's Theorem}

A key fact is that there is a very natural way to choose the branch
cuts. Since the problem is described by an algebraic curve, $V_i(s)$ is
holomorphic with a finite number of branch points. In this case, Stahl's
theorem~\cite{Stahl89,Stahl97}, which proves an earlier conjecture by
Nuttall~\cite{Nuttall77}, states the following:
\begin{enumerate}
\renewcommand{\theenumi}{\alph{enumi}}
\item There exists a unique extremal domain $D$ in which the function has a
  single-valued analytic continuation. In other words, there exists a choice
  of branch cuts $\Gamma$ enforcing single-valuedness such that the resulting
  domain (in our case $\mathbb{C}_\infty\setminus \Gamma$) is the ``biggest''
  one possible (the precise mathematical measure for this is the concept of
  \emph{logarithmic capacity}; this result asserts that there exists a unique
  choice of cuts $\Gamma$ that has minimal logarithmic capacity).
\item The sequence $[L_j/M_j]$ of Pad\'e approximants converges in capacity to
  the function in the extremal domain $D$ for any sequence of indices
  satisfying $L_j+M_j \rightarrow \infty$ and $L_j/M_j \rightarrow 1$ as $j
  \rightarrow \infty$. 
\end{enumerate}
In other words, the near-diagonal Pad\'e approximants of the white germ
converge to the single-valued function in its maximal domain of analytic
continuation, which in this case is everywhere on the $s$-plane except on the
set of branch cuts $\Gamma$. This last step therefore rounds up the method,
providing both a practical way to calculate the analytical continuation and an
unequivocal test for the existence of the operational solution.

\begin{figure}
\includegraphics[width=0.95\columnwidth]{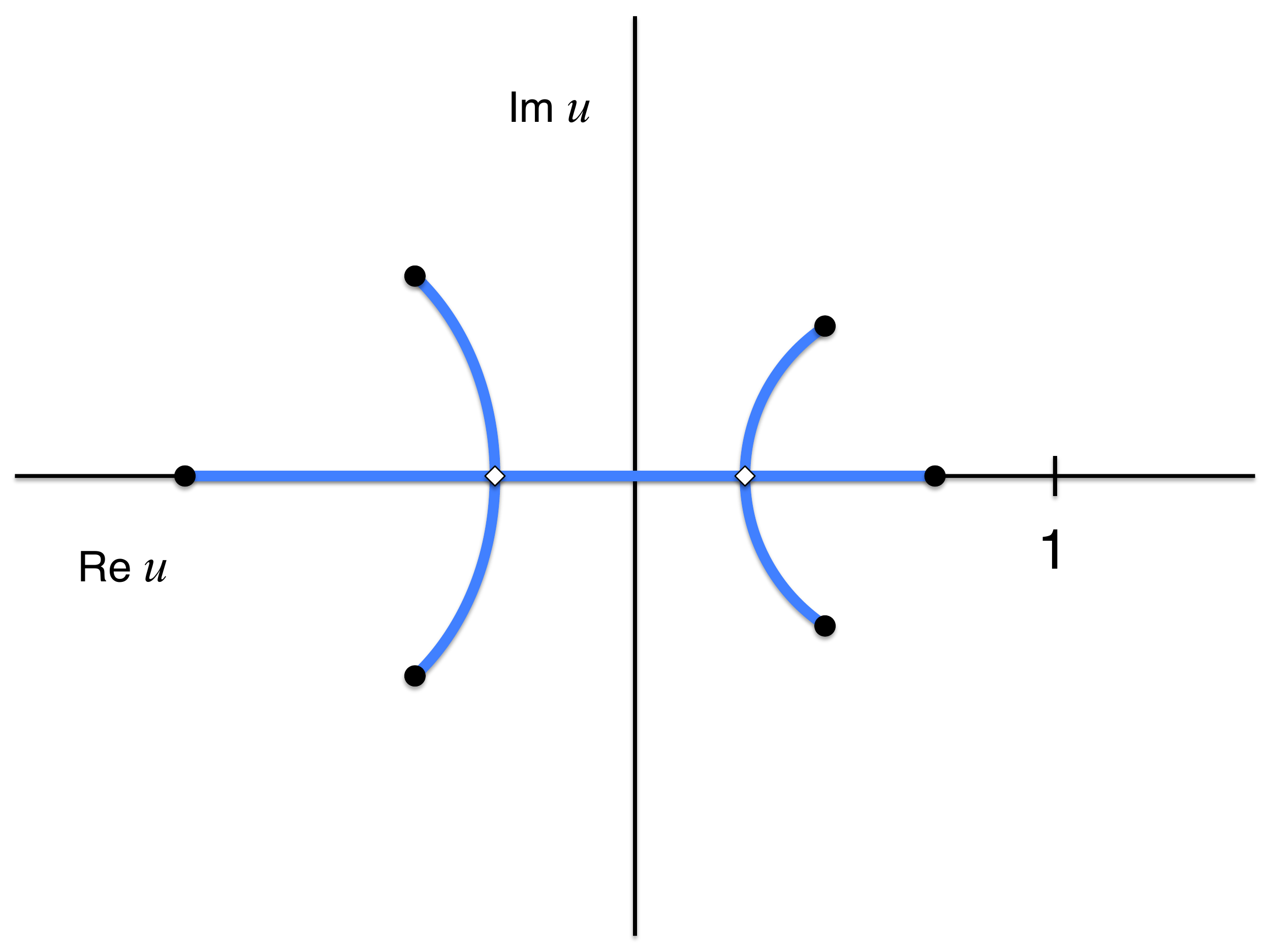}%
\caption{\label{fig:cutset}Schematic example of the minimal cut set $\Gamma$
  in the complex $u$-plane, for an algebraic curve with six branch points
  (black dots). The diamond-shaped points are auxiliary connecting points. In
  this example, the powerflow problem has an operational solution, since the
  path from $u=+\infty$ (i.e. $s=0$) along the real axis can reach $u=1$
  without encountering the cut set, which means that the white germ has an
  analytic continuation along that path. Stahl's theorem proves that the
  analytic continuation can be provided by Pad\'e approximants.}
\end{figure}

Fig.~\ref{fig:cutset} shows a schematic view of a minimal cut set
$\Gamma$. The set has minimal logarithmic capacity in the complex $u$-plane,
where $u=1/s$. For the two-bus problem, the system has only two branch points
and therefore in that case $\Gamma$ is simply a straight segment joining them
(see the Appendix). In the general case, $\Gamma$ would be more
complicated. If one were interested in calculating $\Gamma$, it is interesting
to note that the zeros and poles of the Pad\'e approximants mentioned above
tend to accumulate on this set. In any case, from the point of view of the
powerflow problem, and in particular the existence of an operational solution,
the key question is whether or not the set $\Gamma$ obstructs the path along
the real axis from $s=0$ to $s=1$ (in the $u$-plane, from $u=+\infty$ to
$u=1$).

One subtle point remains: could it ever happen that the excluded set $\Gamma$,
in spite of being of minimal size, crosses or covers the interval $0<s\le1$
when there is otherwise no obstruction to the analytic continuation along that
specific path corresponding to the projective family?  After all, Stahl
guarantees that Pad\'e approximants converge in a maximal domain, but says
nothing about this specific path. The answer lies in the choice of embedding,
as covered in the following section.

Note that the reverse condition is adequately covered: if the operational
powerflow solution does not exist, the procedure will detect so. By our
definition above, if the solution does not exist it means that the white germ
is not analytic-continuable along all values $0<s\le1$ on the real axis, which
is something that can be easily tested by checking the convergence of the
Pad\'e approximants.

\subsection{\label{subsec:embed_choice} Choice of embedding}

Looking back at~\eqref{eq:HolEmb}, it should be remarked that it is certainly
not the only possible embedding satisfying the requirements of the method. For
instance, one could also embed all shunt elements:
\begin{equation}
   \sum_k {Y_{ik}^\text{(tr)} V_k(s)} = - s Y_i^\text{(sh)} V_i(s) 
   + s \frac{S_i^*}{V_i^*(s^*)}
   \label{eq:shuntembed}
\end{equation}
Here $Y_{ik}^\text{(tr)}$ represents the transmission admittances, which
satisfy $\sum_k {Y_{ik}^\text{(tr)}}=0$ for all $i$ (the indices include the
swing bus); while $Y_i^\text{(sh)}$ represent shunt terms, either from load
models or from lines and transformers. The reference state $s=0$ in this case
can be physically interpreted as the line charging susceptances being
completely compensated, waiting for the loads to be connected. This embedding
would yield at $s=0$ a reference solution having all voltages equal to the
swing bus. The authors have used this type of embedding for years and think
that it is probably the most reasonable identity to assign to the empty
network. It will be denoted as HELM's \emph{canonical embedding}.

Another example would be to additionally embed the resistive parts
of $Y^\text{(tr)}$:
\begin{equation}
  j\sum_k {B_{ik}^\text{(tr)} V_k } = - s \sum_k {G_{ik}^\text{(tr)} V_k } 
  - s Y_i^\text{(sh)} V_i + s\frac{S_i^*}{\Vbar_i}
  \label{eq:resistembed}
\end{equation}
so that the resulting linear systems for computing the white germ only involve
the real matrix $B_{ik}^\text{(tr)}$, which could be considered as numerically
more efficient. Yet a further example would be to extend the embedding to the
non-symmetric elements of $B_{ik}^\text{(tr)}$ (those originating from
phase-shifting transformers), so that the final matrix is symmetric and
therefore more efficient to factorize via Cholesky instead of a general LU
decomposition.

However, the choice of embedding is not completely harmless. Different forms
of the embedding result in different algebraic curves, having different
branching points. It can be shown that the introduction of additional embedded
terms in~\eqref{eq:HolEmb}, as suggested above, normally results in
\emph{additional} branch points in the resulting curve. Vast numerical
evidence shows that these additional singularities are harmless in the case
of~\eqref{eq:shuntembed}, since the corresponding Stahl cuts still do not
cover the interval $0<s\le1$ and then the numerical solution is exactly the
same. But in other embeddings, the result may be that, for powerflow cases
where the operational solution at $s=1$ does exist, the additional branching
points could introduce obstacles to the convergence of the Pad\'e
approximants, by way of the minimal set $\Gamma$ covering the point
$s=1$. This is a rather academic problem, since in practice one needs to use
artificial embeddings in order to find a single example of this phenomenon. It
is however an interesting question to find a general way to characterize the
relationship between the functional form of the embedding and the resulting
minimal set $\Gamma$, in order to prevent the use of such problematic
embeddings. This problem is yet unexplored, but it is linked to the
P\'olya-Chebotarev problem, a current research topic in advanced complex
analysis~\cite{Fedorov85,CerdaBharti10}. Nonetheless the authors will report
some considerations about this problem in a forthcoming paper.

For the two-bus case, it can be shown rigorously that the minimal
embedding~\eqref{eq:HolEmb} is free of this problem (see
Appendix~\ref{app:twobus}). For the general $n$-bus case, there is extensive
numerical evidence as well as strong heuristic reasons to support this as
well. The most important one is that the minimal embedding matches the
physically realizable picture of an energized network in which the injections
are progressively turned on, starting from the no-load state. Evidence shows
that the canonical embedding~\eqref{eq:shuntembed} is also free of problems;
in fact it could be argued that the physical picture is even more natural in
this case.

This concludes the foundations for the completeness of the method:
\begin{itemize}
\item If the operational solution exists, the method will find it: Stahl's
  theorem and the choice of the canonical embedding ensure that the Pad\'e
  approximants converge (in capacity measure) to the solution. 
\item If the operational solution does not exist, the method will detect so:
  non-convergence of the Pad\'e approximants (along the path $0<s\le1$ on the
  real axis) necessarily implies an unfeasible powerflow, because if they did
  converge they would be the analytical continuation of the white germ and
  therefore the powerflow solution, thus contradicting the assumption.
\end{itemize}

\section{\label{sec:extensions} Extensions: the role of controls as constraints}

In the context of the powerflow problem, controls are seen as additional
\emph{mathematical constraints} on the voltages or on some other magnitude
(which can always be expressed ultimately in terms of the voltages). Primary
examples are voltage regulation at generator buses, which gives rise to the
so-called PV buses, or on-load tap changing (OLTC) transformers, or switched
capacitor/reactor banks (shunts). Since the powerflow is only concerned with
the steady state, any dynamical aspects are ignored, and thus only the the
stable final state of controls is considered. For instance, in a PV regulated
bus it is only needed to consider the voltage modulus setpoint, and the
mathematical constraint would be simply
$V_kV_k^*=|V_k^\text{sp}|^2$. Obviously, for each new constraint equation
added to the system one needs to free some other parameter in the equation, to
balance the degrees of freedom. In the PV example, one normally designates the
local reactive injection $Q_k$ as the new control variable. In the case of a
remote voltage-controlled PQ bus $k$, the new variable could be some other
$Q_{k'}$ or some weighted combination of reactive injections.

The HELM method can seamlessly incorporate any type of control, as long as the
associated constraints can be expressed as some algebraic function of the
voltages or flows. This is always the case since all controls can be expressed
as a polynomial equation in the steady state. These will be referred to as
algebraic controls. This way the system is still described by an algebraic
curve and thus all the fundamental properties of the method are preserved. In
practice this means that all ``smooth'' controls can be accommodated in HELM,
but there are two characteristics that break holomorphicity and therefore need
to be treated outside of HELM methodology: saturation limits and discreteness.
Real controls always exhibit resource limits, and sometimes also discreteness
effects. The most prominent examples of saturation limits are the
$(Q_\text{min}, Q_\text{max})$ capability limits in generator buses and the
tap ratio range limits in OLTC transformers. The latter are also prime
examples of discreteness (tap ratio steps) and threshold activation effects
(control deadband). All of these issues can be dealt with via successive
application of the algebraic method, combined with discretization techniques
such as relaxation and/or combinatorial search. Further details and a complete
framework for dealing with limits and discreteness will be the subject of a
forthcoming publication. The exposition here will only deal with algebraic
controls, focusing on the HELM fundamentals.

\subsection{Algebraic controls}

The introduction of controls in the HELM framework is guided by the
fundamentals exposed in the previous section. It basically amounts to an
adequate embedding of the constraint expression, in a way that preserves the
algebraic nature of the problem (and therefore holomorphicity) and also
attains a reference solution at $s=0$ that makes physical sense as an
energized, no load network. On the other hand, for the addition of each new
constraint one has to designate one new free variable, in order to keep the
number of equations and unknowns balanced. This is best exemplified by showing
how this can be done in the simplest and most ubiquitous case, the PV node.

Let us consider the general powerflow equations of a network as
in~\eqref{eq:LF}, except that now the buses are labeled into two sets: the
$\{PV\}$ set, having $\Npv$ buses, and the $\{PQ\}$ set having $\Npq$. There
are $\Npv$ new constraints $V_kV_k^*=|V_k^\text{(sp)}|^2$, which are clearly
algebraic. In a PV bus, voltage control is achieved by regulating the
injection of reactive power local to the bus. This naturally suggests the new
variables of the system, needed to counterbalance the constraints.

These constraints need to be holomorphically embedded using the simplest
functional form possible in order to avoid introducing extra singularities in
the $s$-plane. In this case one cannot simply use a constant function
$V_k(s)V_k^*(s^*)=|V_k^\text{(sp)}|^2$, since the voltages at $s=0$ could not
satisfy it without some reactive injection, and these are switched off at the
reference state. The simplest valid form is a \emph{linear} function
interpolating between the natural values of voltage at $s=0$ and the desired
setpoint at $s=1$. Assuming without loss of generality that the swing has a
unit voltage setting, the embedding is simply:
\begin{equation}
  V_k(s) V_k^*(s^*) = 1 + s \left( |V_k^\text{(sp)}|^2 - 1 \right)
  \label{eq:PVconstraint_embedded}
\end{equation}
Separating nodes into the PQ set and the PV set, this is the proposed
canonical embedding:
\begin{equation}
  \begin{split}
    \sum_l Y_{kl}^\text{(tr)} V_l(s) &= - s Y_k^\text{(sh)} V_k(s)
    + s\frac{P_k}{V^*_k(s^*)} - j\frac{Q_k(s)}{V^*_k(s^*)} \\
    \sum_j Y_{ij}^\text{(tr)} V_j(s) &= - s Y_i^\text{(sh)} V_i(s)
    + s\frac{S^*_i}{V^*_i(s^*)}
  \end{split}
  \label{eq:PVembed_canonical}
\end{equation}
where the indices $k\in\{PV\}$ and $i\in\{PQ\}$, and the rest go over all
buses unless otherwise noted. The enlarged
system~\eqref{eq:PVconstraint_embedded}+\eqref{eq:PVembed_canonical} preserves
all algebraic properties of the original, so that all of HELM theory and
methods apply. The variables $\Vbar_i$ would be introduced analogously, and
the new system would then consist of $3n_{PV}+2n_{PQ}$ equations:
$2(n_{PV}+n_{PQ})$ equations coming from~\eqref{eq:PVembed_canonical} and its
reflection, plus $n_{PV}$ new equations coming
from~\eqref{eq:PVconstraint_embedded}, whose reflection yields identically the
same expression. Correspondingly, there are $3n_{PV}+2n_{PQ}$ variables:
$2(n_{PV}+n_{PQ})$ variables $Vi, \Vbar_i$, plus $n_{PV}$ new variables $Q_k$.

Moreover this embedding allows a meaningful reference state at $s=0$, namely,
the zero injection state. Note that this requires that the new variables
$Q_k(s)$ be zero at $s=0$. The voltages then become all equal to the swing bus
voltage at $s=0$.

The numerical procedure described in~\cite{Trias12} would then be applied
analogously. Using the power series expansion for the variables
$V(s)=\sum_N V[N] s^N$ and $Q(s) =\sum_N Q[N] s^N$, and substituting into
equations~\eqref{eq:PVconstraint_embedded}+\eqref{eq:PVembed_canonical}, one
would obtain a linear system where the coefficient matrix is fixed, and the
terms at order $N$ can be obtained from the results computed at the preceding
orders. Solving this system sequentially in successive orders and computing
the Pad\'e approximants as shown in~\cite{Trias12} would conclude the method.

However, the fact that the system is converted in this fashion into a sequence
of linear systems provides the chance to perform some powerful
simplifications, simply following standard transformations akin to Gaussian
elimination. For instance, in this case it can be shown that the linear system
can be reduced from dimension $3n_{PV}+2n_{PQ}$ to $n_{PV}+2n_{PQ}$. Further
details on the computationally efficient implementation of this control and
several others will be the subject of a future publication.

This concludes the method for incorporating PV nodes. It should be remarked
that other types of controls, even sophisticated ones involving several
control variables and controlled magnitudes (e.g., area interexchange
schedules) can be integrated in HELM analogously in an exact way.

\section{Conclusion}

The HELM method and its associated theory opens up new perspectives on the
powerflow problem. It sheds new light on the problems of existence,
multiplicity of solutions, and voltage collapse. It achieves this through the
use of advanced concepts in algebraic geometry and complex analysis. Through
the study of a fundamental projective invariance, it is shown how the original
problem transforms into the study of an algebraic curve. Then recent advances
in the field of rational approximants provide a practical way to compute
solutions through power series expansions, and additionally confer
completeness to the method by virtue of a theorem on the maximality of the
analytic continuation provided by the Pad\'e approximants.

\section*{Acknowledgments}
The author would like to thank J. L. Mar\'in for his outstanding contribution
in the clarification of many concepts basic to this paper.

\Appendix
\section{\label{app:twobus} Full HELM-based solution of the two-bus problem}

The two-bus model provides an easy and rather complete showcase for all the
elements of HELM theory. It allows exemplifying all the fundamentals exposed
in the preceding sections through explicit, closed form calculations. Let us
start with the powerflow equations written in dimensionless magnitudes
$U\equiv\frac{V}{V_0}$ and $\sigma\equiv\frac{ZS^*}{|V_0|^2}$:
\begin{equation}
  U = 1 + \frac{\sigma}{U^*} 
  \label{eq:twobus}
\end{equation}
Solving separately for the real and imaginary parts of $U$, it is
straightforward to arrive at the exact solution:
\begin{equation}
  U = \frac{1}{2} \pm \sqrt{\frac{1}{4}+\sigma_R-\sigma_I^2} + j\sigma_I
  \label{eq:twobussoln}
\end{equation}
subject to the condition $\Delta\equiv\frac{1}{4}+\sigma_R-\sigma_I^2\geq 0$,
where $\sigma_R, \sigma_I$ are the real and imaginary parts of $\sigma$,
respectively. It is important to realize that if this condition is not met,
then there is no solution to the powerflow problem. Therefore as a parametric
ordinary equation the system has the two solutions~(\ref{eq:twobussoln}) if
$\Delta\geq 0$, and none if $\Delta<0$.

Now consider the corresponding embedded system, written in explicit polynomial
form:
\begin{equation}
  \begin{split}
    U(s) \Ubar(s) - \Ubar(s) - s\sigma &= 0 \\
    \Ubar(s) U(s) - U(s) - s\sigma^* &= 0
  \end{split}
  \label{eq:twobuseembedding}
\end{equation}
In this case it is straightforward to carry out the elimination procedure by
hand. Eliminating $\Ubar$ from the second equation and substituting in the
first, one readily arrives to the algebraic curve:
\begin{equation}
  \mathfrak{P}(s,U) = U^2 - (1 + 2js \sigma_I)\, U - s \sigma^* = 0
  \label{eq:twobuspoly}
\end{equation}
and the other variable can be obtained in terms of $U$, also in polynomial form, by
eliminating $U\Ubar$ from~\eqref{eq:twobuseembedding}:
\begin{equation*}
  \Ubar = U - 2js\sigma_I
\end{equation*}

Solving~\eqref{eq:twobuspoly} one readily obtains the two branches of this
curve, together with the corresponding ones for $\Ubar(s)$:
\begin{equation}
  \begin{split}
    U(s)_\pm &= \frac{1}{2} \pm \sqrt{\frac{1}{4}+s\sigma_R-s^2\sigma_I^2} + js\sigma_I \\
    \Ubar(s)_\pm &= \frac{1}{2} \pm \sqrt{\frac{1}{4}+s\sigma_R-s^2\sigma_I^2} - js\sigma_I
  \end{split}
  \label{eq:twobusbranches}
\end{equation}
The branches exist everywhere on the complex $s$-plane (and coincide at
the two branch points). However, for them to be a solution of the powerflow
problem, the reflection condition $\Ubar(s)=U^*(s^*)$ has to be satisfied at
$s=1$. Additionally, the branch containing the operational solution, if it
exists, is easily identified as the one with the plus sign
in~(\ref{eq:twobusbranches}).

One verifies that the reflection condition is satisfied everywhere on the
$s$-plane except at the points where the discriminant
$\Delta(s)\equiv\frac{1}{4}+s\sigma_R-s^2\sigma_I^2$ becomes real negative (in
this case, this holds for both branches). Solving for the roots of this
discriminant,
\begin{equation}
  s_\pm = \frac{\sigma_R \pm |\sigma|}{2\sigma_I^2}
  \label{eq:twobusbranchpoints}
\end{equation}
we find that the reflection condition fails for values $s<\sminus$ and
$s>\splus$ on the real axis. Note that the points $s_\pm$
in~\eqref{eq:twobusbranchpoints} are the \emph{branch points} of the algebraic
curves~\eqref{eq:twobusbranches}, i.e.\ the points where the $(+)$ and $(-)$
branches collide. The value $\sminus$ is always negative; since we are
interested in reaching $s=1$, the condition for existence of powerflow
solutions is therefore $\splus\geq 1$. From~\eqref{eq:twobusbranchpoints}
above, this translates to $\frac{1}{4}+\sigma_R-\sigma_I^2\geq 0$, thus
recovering the condition found in~\eqref{eq:twobussoln}.

On the other hand, these intervals $s<\sminus$ and $s>\splus$ form a simply
connected set of \emph{branch cuts} of the algebraic curve, as they join the
two branch points through infinity (or equivalently, through $0$ in the
$1/s$-plane). Since the arc of minimal logarithmic capacity joining two points
is always a straight segment, this proves that this is the minimal cut-set in
the sense of Stahl. Therefore the Pad\'e approximants, for both germs, will
fail to converge for $s<\sminus$ and $s>\splus$ on the real axis. This
confirms the general theory: the sequence of near-diagonal Pad\'e approximants
converge to the solution when it exists, and they do not converge when it does
not.

\subsection{The PV case}

The two-bus PV case solves analogously, with the addition of an embedded
constraint as in~\eqref{eq:PVconstraint_embedded} and the corresponding new
variable $Q$.  It is also straightforward to carry out the elimination
procedure by hand, but it is interesting to see how Buchberger's
algorithm~\cite{Buchberger98} would do it.  Writing the constraint as
$K(s)\equiv 1+s(|V^\text{sp}|^2-1)$, and choosing lexicographic order
$U\prec Q \prec \Ubar$, the system is now written as the following polynomial
ideal in the ring of polynomials of three variables:
\begin{equation*}
  \left<
    \Ubar U-\Ubar+jszQ-szP, \quad
    \Ubar U-jsz^*Q-U-sz^*P, \quad \Ubar U - K(s)
  \right>
\end{equation*}
where we have defined $z\equiv\frac{Z}{|V_0|^2}\equiv r+jx$. The Gr\"obner
elimination procedure is simplified in this case if one assumes a lossless
line, $r=0$:
\begin{equation*}
  \left<
    \Ubar U-\Ubar-sxQ-jsxP, \quad
    \Ubar U-sxQ-U+jsxP, \quad \Ubar U - K(s)
  \right>
\end{equation*}
Elimination of the leading monomial $\Ubar U$ by subtracting the first two
polynomials, plus using the third one, leads to:
\begin{equation*}
  \left<
    \Ubar U - K(s), \quad
    \Ubar-U+2jsxP, \quad sxQ + U -jsxP - K(s)
  \right>
\end{equation*}
To eliminate the leading monomial, multiply the second polynomial by $U$ and
subtract. This yields the Gr\"obner basis in triangular form:
\begin{equation*}
  \left<
    \Ubar-U+2jsxP, \quad
    sxQ + U -jsxP -K(s), \quad U^2 -2jxsPU - K(s)
  \right>
\end{equation*}
All solutions are therefore completely determined by the solutions to $U$ from
the last polynomial in the basis:
\begin{equation*}
  U = jxsP \pm \sqrt{K(s)-x^2s^2P^2},
\end{equation*}
since $Q$ and $\Ubar$ are given by the second and first polynomials of
the basis, respectively. The operational branch is clearly the one with the
plus sign, and the feasibility condition in the case is $K(s)-x^2s^2P^2\ge0$.


\end{document}